\documentclass[preprint]{revtex4-1}

\usepackage{graphicx}
\usepackage{caption}
\usepackage{subcaption}

\begin{document}
\title{Absolute measurement of the ${}^{1}S_{0}$ -- ${}^{3}P_{0}$  clock transition in neutral ${}^{88}$Sr over the 330~km-long stabilized fibre optic link}


\author{Piotr Morzy\'nski$^1$, Marcin Bober$^1$, Dobros\l{}awa Bartoszek-Bober$^1$,    Jerzy Nawrocki$^3$,     Przemys\l{}aw Krehlik$^2$, \L{}ukasz \'Sliwczy\'nski$^2$, Marcin Lipi\'nski$^2$, Piotr Mas\l{}owski$^1$, Agata Cygan$^1$,  Piotr Dunst$^3$, Micha\l{} Garus$^1$, Daniel Lisak$^1$, Jerzy Zachorowski$^4$, Wojciech Gawlik$^4$, Czes\l{}aw Radzewicz$^5$, Roman Ciury\l{}o$^1$, Micha\l{} Zawada$^{1,*}$}

\address{$^1$Institute of Physics, Faculty of Physics, Astronomy and Informatics, Nicolaus Copernicus University, Grudzi\c{a}dzka 5, PL-87-100 Toru\'n, Poland\\
$^2$Department of Electronics, AGH University of Science and Technology, al. Mickiewicza 30, PL-30-059, Krak\'ow, Poland \\
$^3$Time and Frequency Department, Astrogeodynamic Observatory of Space Research Center, Borowiec, Drapa\l{}ka 4, PL-62-035 K\'ornik,  Poland\\
$^4$M. Smoluchowski Institute of Physics, Faculty of Physics, Astronomy and Applied Computer Science, Jagiellonian University, St. \L{}ojasiewicza 11, PL-30-348 Krak\'ow, Poland\\
$^5$Institute of Experimental Physics, Faculty of Physics, University of Warsaw, Pasteura 5, PL-02-093 Warsaw, Poland}

\email{zawada@fizyka.umk.pl} 

\begin{abstract}
We report a stability below $7\times 10{}^{-17}$ of two independent optical lattice clocks operating with bosonic ${}^{88}$Sr isotope. The value (429~228~066~418~008.3(1.9)${}_{syst}$(0.9)${}_{stat}$~Hz) of the absolute frequency of the ${}^{1}S_{0}$ -- ${}^{3}P_{0}$ transition was measured with an optical frequency comb referenced to the local representation of the UTC by the 330 km-long stabilized fibre optical link. The result was verified by series of measurements on two independent optical lattice clocks and agrees with recommendation of Bureau International des Poids et Mesures.
\end{abstract}


\flushbottom
\maketitle
%
%
\thispagestyle{empty}

\flushbottom
\section*{Introduction}
Ultracold neutral atoms in an optical lattice~\cite{Ido03} are seen as an alternative to single-ions~\cite{Rosenband08} for development of optical frequency standards. All best present realizations of the strontium optical clocks are made with fermionic strontium isotope ${}^{87}$Sr~\cite{Ye14, LeTargat13, Hinkley13, Falke14, Ushijima14}, since the bosonic isotopes are expected to have larger collisional effects on the clock transition. Additionally, the bosonic isotopes require at least one extra field to induce the clock transition, which implies careful control of this field and its respective field shift. On the other hand, the bosonic lattice clocks have some advantages over their fermionic counterpart: no first order Zeeman shift, no vector or tensor lattice Stark shifts and much higher isotopic abundance.  Lack of  hyperfine structures in both ${}^{1}S_{0}$ and ${}^{3}P_{0}$ states and higher abundance reduce the time required for one lock cycle. Furthermore, the set-up of cooling and trapping the bosonic isotope is simpler, which is important for transportable systems.

The experimental difficulties in limiting and characterising the collisional shift in bosons are the reason why there are only two reported  measurements  of the ${}^{1}S_{0}$ -- ${}^{3}P_{0}$  transition in ${}^{88}$Sr so far~\cite{Baillard07,Akatsuka08}.
 To calculate the recommended frequency values for the practical realizations of the metre (MeP) and secondary representations of the second (SRS), the BIPM takes into account the weighted average of independently obtained frequencies.
  A limited pool of available measurements forced the BIMP to set practical relative uncertainties above the $1\times 10^{-14}$ level when the ${}^{1}S_{0}$ -- ${}^{3}P_{0}$  transition in ${}^{88}$Sr is used as  MeP and restrain in recommending this transition as  SRS~\cite{BIPM88}. 
  
 There are  two known ways to limit the effects of the collisions: the first, the measurements in the optical lattice trap with low atomic density and high confinement to suppress tunnelling effects \cite{Lisdat09}; the second, the use of higher dimensional optical lattice trap~\cite{Akatsuka08}. In our system the low value of collisional shift is ensured by a large waist of the lattice and trapping only a few atoms per lattice site in a trap.
 We report a system of two independent bosonic strontium optical lattice standards with ${}^{88}$Sr probed with a single shared ultranarrow
laser. The absolute frequency of the clock transition is measured by the use of a frequency-doubled  Er:fibre polarization-mode-locked optical frequency comb referenced 
 to the UTC(AOS) and UTC(PL) \cite{Azoubib03,Jiang15} via the 330~km-long  stabilized fibre optic link of the OPTIME network~\cite{Sliwczynski13,Krehlik15}.

\section*{Methods}

\subsection*{Optical Lattice Standards}

The experimental set-up of our system has been described in detail in Ref.\cite{Bober15}, so only its most essential elements are presented below.

\begin{figure}[h!]
\centering
\includegraphics[width=0.5\columnwidth]{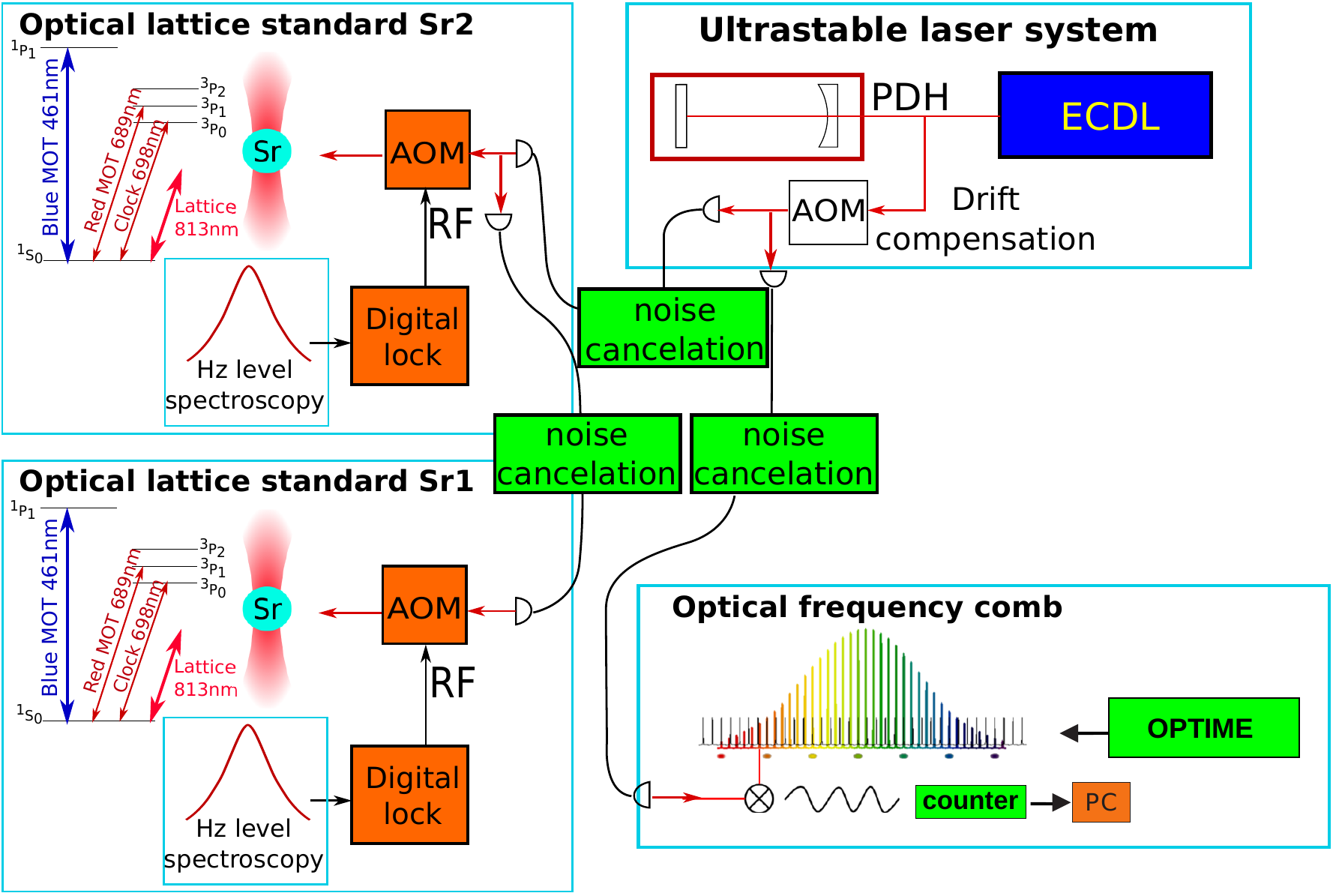}
\caption{  {\bf A simplified scheme of the system of two optical lattice clocks Sr1 and Sr2.}  The clouds of atoms in Sr1 and Sr2 are independently probed by two beams from an ultrastable laser.  The frequencies of both beams are locked to the narrow resonances in each standard by a digital lock and acousto-optic frequency shifters (AOM). The frequencies of each clock transitions can be compared by the use of an optical frequency comb to the UTC(AOS) and UTC(PL) \cite{Azoubib03,Jiang15} via the 330~km-long  stabilized fibre optic link of the OPTIME network~\cite{Sliwczynski13,Krehlik15}.
 \label{fig:general}}
\end{figure}

A simplified scheme of the system of two optical lattice clocks is depicted in Fig. \ref{fig:general}.
Two optical frequency standards (Sr1 and Sr2) are based on the  ${}^{1}S_{0}$ -- ${}^{3}P_{0}$ transition in neutral ${}^{88}$Sr atoms.
Two clouds of cold atoms in Sr1 and Sr2, trapped in the vertical optical lattices, are independently probed by an ultrastable laser with spectral width below 1~Hz. 
The laser beam is split into two optical paths.
The frequencies of both beams are independently digitally locked to the narrow atomic resonances in each standard by  feedback to the  acousto-optic frequency shifters.

The short-time frequency reference of the optical standards, i.e. the ultrastable laser, is an Extended Cavity Diode Laser (ECDL)  locked to the TEM${}_{00}$ mode of the high-Q cavity. The light from the ultrastable laser is transferred to the Sr1 and Sr2 standards and to the optical frequency
comb through optical fibres. Each fibre has a system of active Doppler cancellation of the fibre-link noises to
assure the transfer of stable optical frequencies \cite{Ma94}.

In both Sr1 and Sr2 systems the Fabry-Perot diode lasers are injection-locked to the light from ultrastable laser. The master-slave system filters out
any power fluctuations of the injection laser. 
The beam is passing the acousto-optic modulator (AOM) of the digital lock and is injected to the optical lattice such that it is exactly superimposed with the lattice. The beam waist is much bigger than the size of the sample of atoms.

\subsection*{Stabilized fibre optic link and UTC(AOS)}

The frequencies of the clock transitions can be compared by the use of an optical frequency comb with the UTC(AOS) and UTC(PL)\cite{Azoubib03,Jiang15} via the OPTIME network \cite{Sliwczynski13}.

\begin{figure}[h!]
\includegraphics[width=\textwidth]{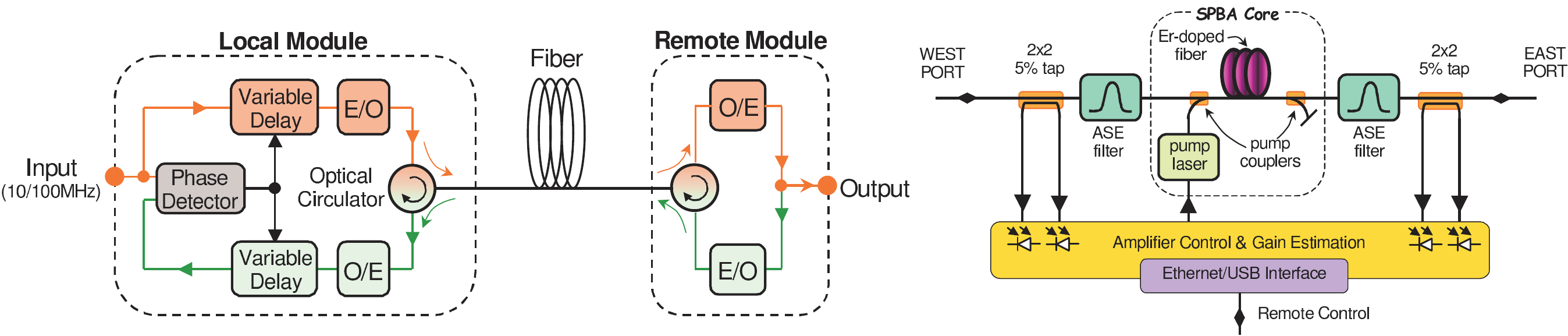}
 \caption{{ {\bf The ELSTAB system.} Left panel: a simplified block diagram of the local and remote modules in the ELSTAB system. E/O and O/E denote the electro-optic and optical-electric converters, respectively. Right panel: single-path bidirectional amplifier (SPBA) diagram. }
 \label{fig:elstab}}
\end{figure}

The  330~km-long time  and frequency dissemination line between the Space Research Centre at Borowiec Astrogeodynamic Observatory (AOS) and KL FAMO in Toru\'n is electronically stabilized with the ELSTAB technology \cite{Sliwczynski11}. 
The underlying idea of the ELSTAB solution is to implement the compensation of the fibre delay fluctuations in the electronic domain, by using a pair of precisely matched variable delay lines. The delay lines are both placed  in the forward and backward paths of the delay-locked-loop (DLL) structure (see Fig.~\ref{fig:elstab}, left panel).

 The local module is installed at the AOS in Borowiec and the remote module is installed at the KL FAMO in Toru\'n.  Additionally, the line contains seven specialized optical bidirectional amplifiers based on erbium-doped fibres  (see Fig.~\ref{fig:elstab}, right panel). Thanks to bidirectional operation over the same optical path for the forward and backward
directions, the propagation delay is constant for both directions. Consequently, the possible phase fluctuations compensate and the insertion of the amplifier does 
not destroy the symmetry of the optical path.

\begin{figure}[h!]
\centering
\includegraphics[width=0.5\columnwidth]{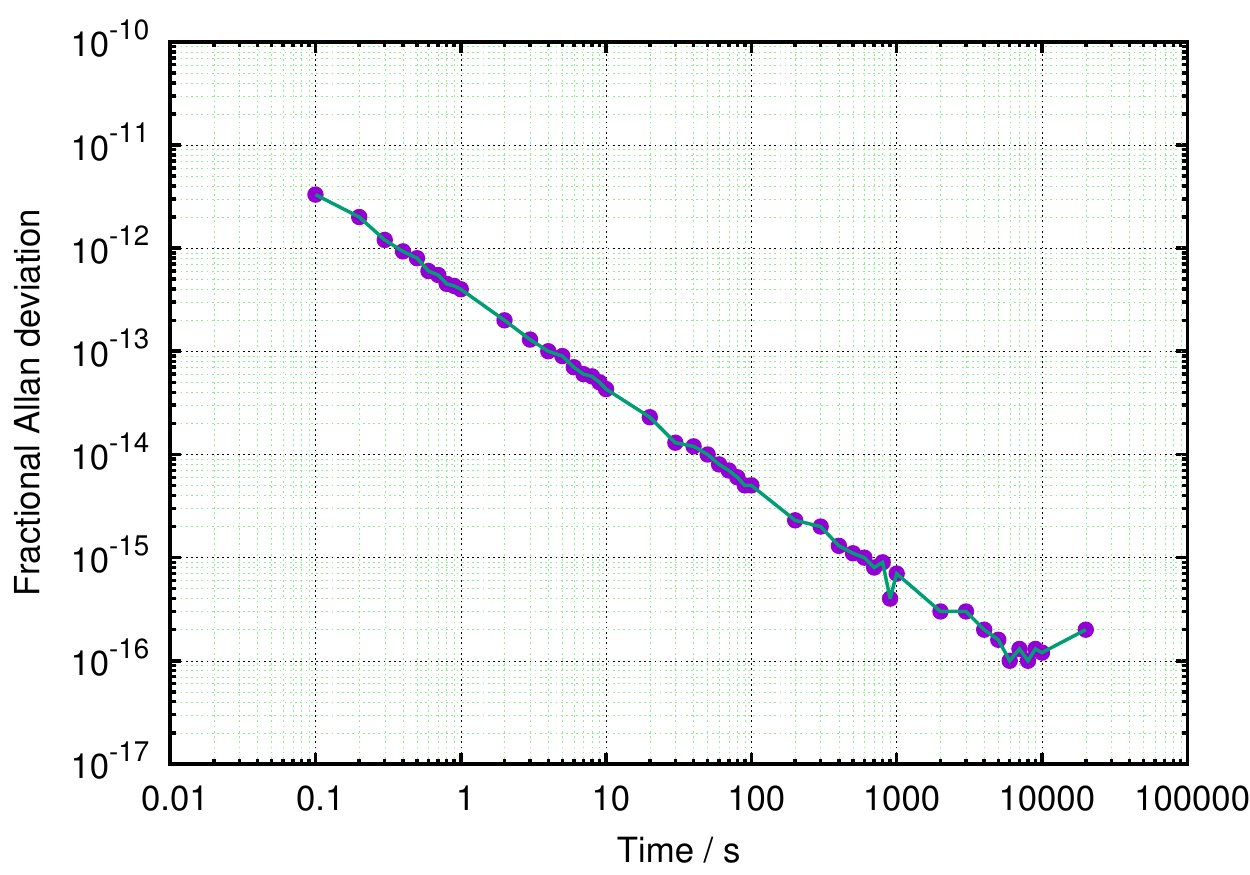}
\caption{ { {\bf The quality of the fibre link.}} Frequency transfer stability obtained during the pre-installation test of the ELSTAB system in fractional units represented by the Allan standard deviation.
 \label{fig:bounce}}
\end{figure}

To estimate the quality of the link, the pre-installation tests with a 300~km-long fiber on spools and bidirectional optical amplifiers were performed. The stability of the remote 10~MHz signal was measured with respect to the local input, using the A7-MX Signal Stability Analyser. The overlapping Allan deviation is equal to $4\times 10^{-13}$ for 1~s integration period, and drops down to $3\times 10^{-16}$ within 1~h (Fig. \ref{fig:bounce}).

\begin{figure}[h!]
\centering
\includegraphics[width=0.6\columnwidth]{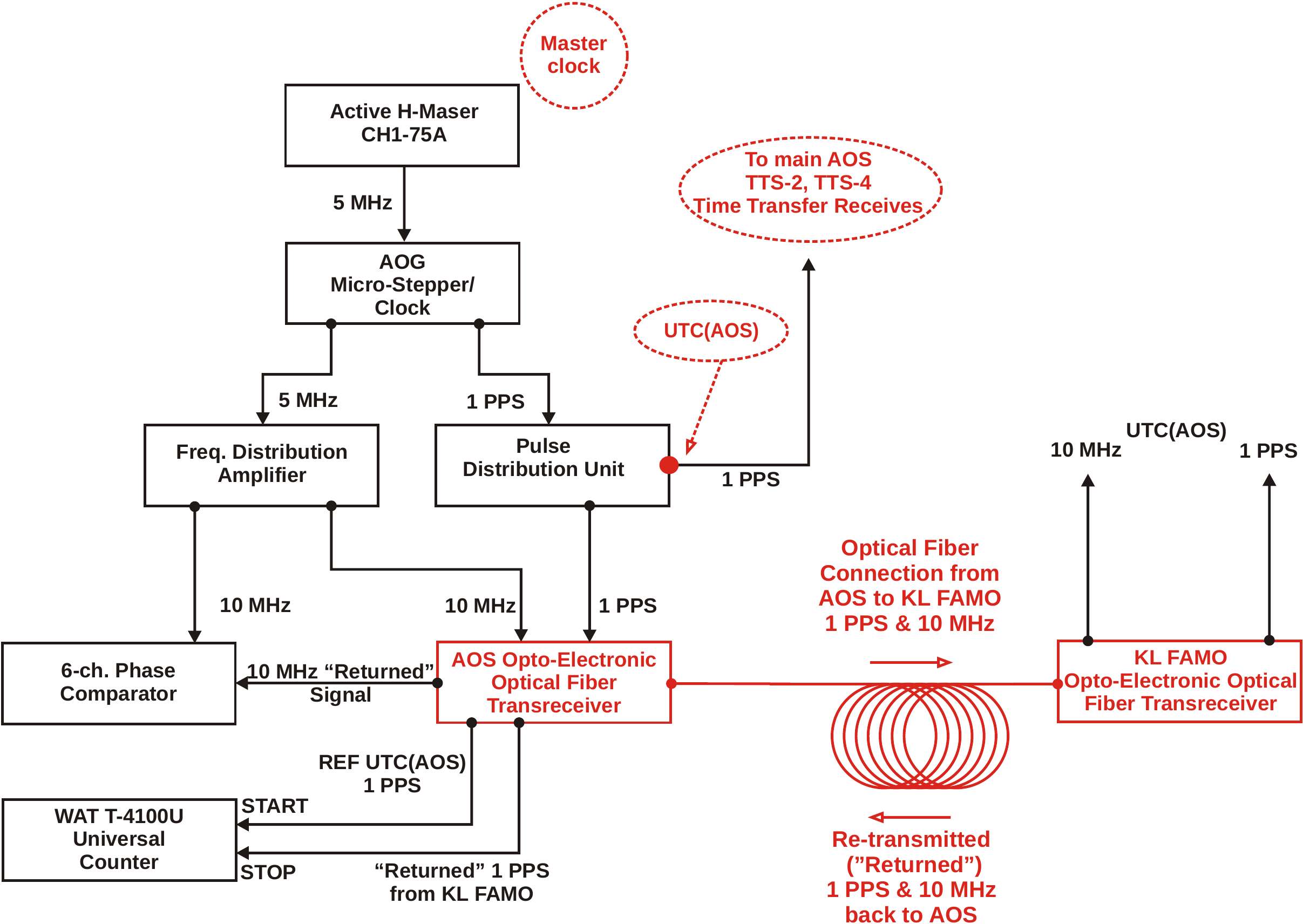}
\caption{ { {\bf The frequency chain at the AOS in Borowiec.} The local representation of the Coordinated Universal Time at AOS in 
Borowiec, UTC(AOS), is realized directly in the form of a 1PPS (one-pulse-per-second) by  
a system of an active H-maser and an offset generator (AOG).  The AOG compensates 
the linear frequency drift of the maser on a daily basis and adds corrections in respect to 
the UTC, extrapolated from differences UTC- UTC(AOS) and UTCr-UTC(AOS) published  
monthly and weekly, respectively, in Circular-T~\cite{CircT}.}
 \label{fig:aos}}
\end{figure}

The local representation of the Coordinated Universal Time (see e.g. \cite{Whibberley11}) at AOS in Borowiec, UTC(AOS), is realized directly in the form of a 1PPS (one-pulse-per-second) by a system of an active H-maser (CH1-75A) and an offset generator (Symmetricom  Auxiliary Output Generator -- AOG). The active H-maser provides good stability over measurement times of up to a few days, with an Allan deviation of  $2 \times 10^{-13}$ at an averaging time of $\tau=1$~s and decreasing as $1/\sqrt\tau$ up to the averaging time of $\tau=10^{4}$~s.  The AOG compensates the linear frequency drift of the maser on a daily basis and adds corrections in respect to the UTC, extrapolated from differences UTC- UTC(AOS) and UTCr-UTC(AOS) published monthly and weekly, respectively, in Circular-T \cite{CircT}. The details of the frequency chain at the AOS are presented in Fig.\ref{fig:aos}.

\section*{Results}

\subsection*{Statistical stability of the system}

\begin{figure}[h!]
\centering
\includegraphics[width=0.5\columnwidth]{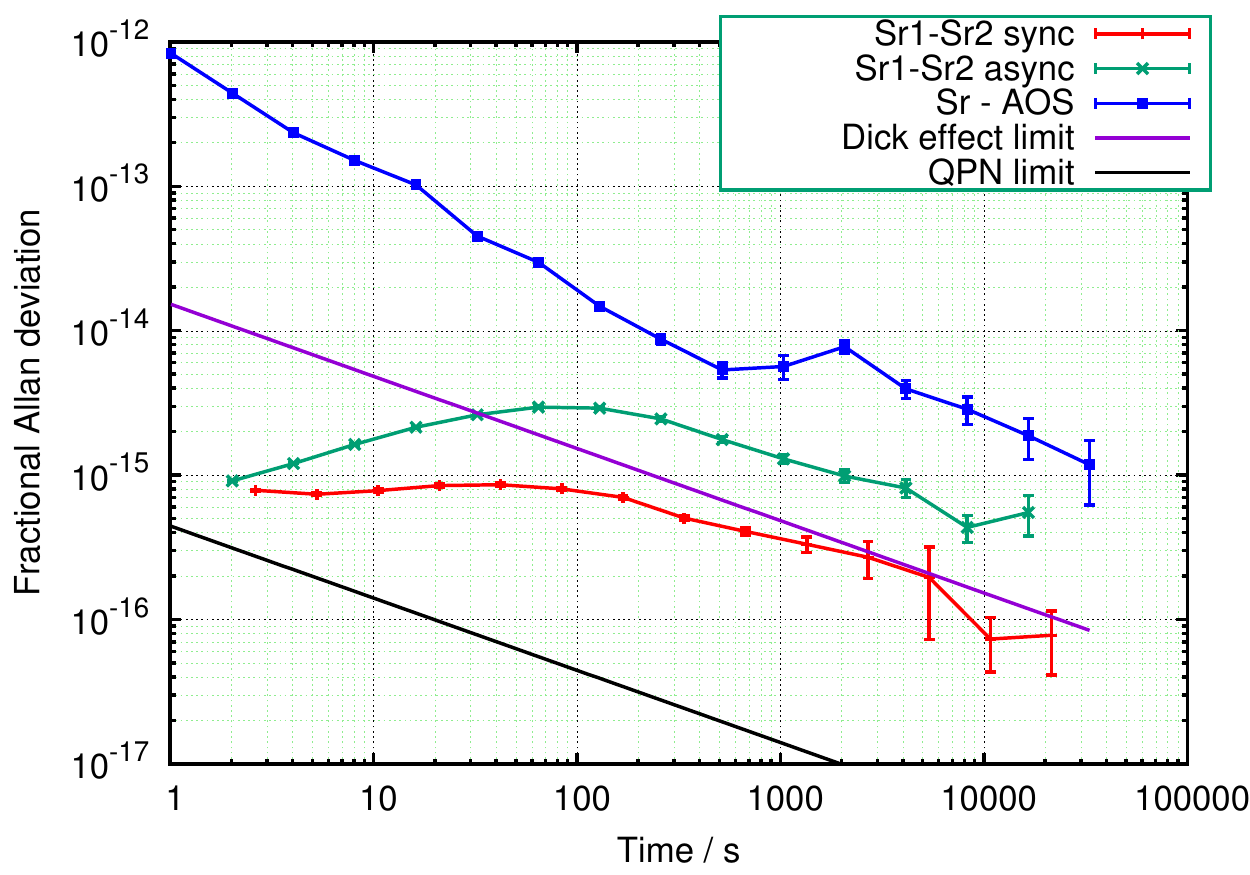}
\caption{{{\bf  The measured frequency stability.} The frequency difference of the two synchronously and asynchronously operated
two optical lattice standards (red and green lines, respectively)} and between Sr1 system and the UTC(AOS) (blue { line}) in fractional units represented by the Allan standard deviation. The comparison between Sr1 and UTC(AOS) was made over the  dedicated  330~km-long stabilized fibre optic link. { The two fundamental limitations of the optical clock stability, i.e. the quantum projection noise (QPN) limit, which is $\sqrt{2} $ times the QPN limit for N = 32000 atoms of an individual clock,
and the Dick effect limit, derived from the power spectral density of the ultrastable clock laser for a clock cycle time of 1.32~s,  are depicted by black and violet lines, respectively.}
 \label{fig:allan}}
\end{figure}

The difference between the corrections of the digital locks in the Sr1 and Sr2 standards yields the momentary frequency difference between the two clocks. The measured frequency stability in fractional units represented by the Allan standard deviation is presented in Fig.~\ref{fig:allan} { with two standards  operating  synchronously and asynchronously (red and green lines, respectively). The short-time stability of the standards, up to 100~s of averaging, is limited by the quality of the ultrastable laser. The synchronous operation is mostly free from ultrastable laser fluctuations, except of small residual caused by independent setting of digital locks in the two standards. The measured stability of the synchronously operated clocks, with a clock cycle time of 1.32~s and interrogation time of 40~ms, reached $7\times 10^{-17}$ after 10000~s of averaging.  The two fundamental limitations of the optical clock stability, i.e. the quantum projection noise (QPN) limit\cite{Santarelli99} for N = 32000 atoms of an individual clock 
and the Dick effect limit\cite{Dick87,Santarelli98}, derived from the power spectral density of the ultrastable clock laser and the clock cycle time, are also depicted in Fig.~\ref{fig:allan}.
}

The stability of the Sr1 was also compared with the stability of the UTC(AOS) maintained by the hydrogen maser in AOS in Borowiec. The comparison was made over the dedicated  330~km-long stabilized fibre optic link ({ blue line} in Fig.~\ref{fig:allan}). This measured frequency stability provides information about the overall statistical uncertainty of the reference frequency of the hydrogen maser, the stability of the fibre link, and of optical frequency comb. { For example, the plateau at around 2000~s corresponds to the  $\pm 1^\circ$C fluctuations of temperature in the room where part of the frequency chain at the AOS in Borowiec (micro-stepper and frequency distribution amplifiers) is placed. }

\subsection*{Accuracy budget}

\begin{figure}[h!]
\centering
\includegraphics[width=\textwidth]{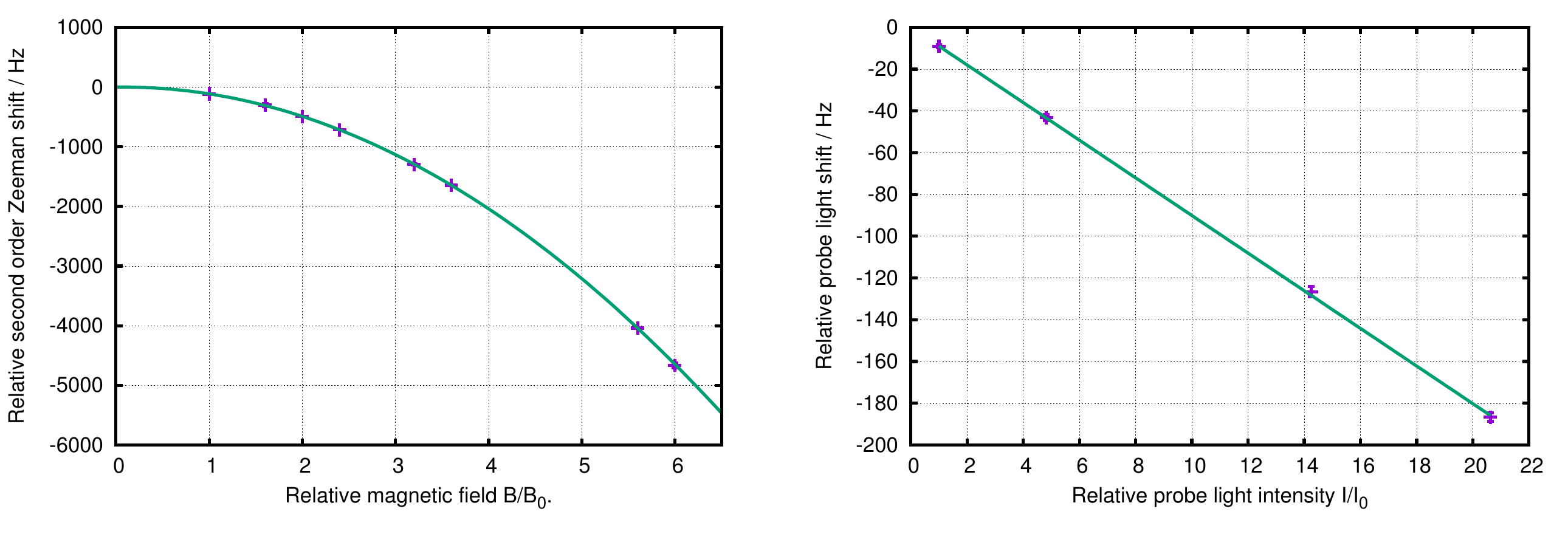}
\caption{ {\bf Examples of evaluations  of systematic shifts.} 
Left panel: evaluation of the quadratic Zeeman shift { in Sr2}. The second order Zeeman correction depends on the absolute value of the magnetic field therefore similar measurements are made in all three directions  { B${}_0$ = 2.383 mT corresponds to the applied B-field inducing clock transition at standard operating conditions. Right panel: evaluation of the probe light shift in Sr2}. 
 { I${}_0$ = 488 mW/cm${}^2$ corresponds to the clock laser intensity at standard operating conditions. }
 \label{fig:shifts}}
\end{figure}

We have evaluated the main contributions to the frequency shifts in both Sr1 and Sr2 standards.{ The accuracy budgets are compared  in Table \ref{tab:acc} for typical experimental conditions: applied B-fields inducing clock transition equal to 2.725 and 2.383 mT,  clock laser intensities equal to 207 and 488 mW/cm${}^2$ and  resultant Rabi frequencies\cite{Taichenachev06} equal to 7.4 and 9.7 Hz for Sr1 and Sr2, respectively.} 
Most of the systematic contributions presented there were evaluated by making a series of several simultaneous (interlaced) locks to the atomic line with different values of particular physical parameter in one of the standard, with the other standard serving as a stable reference. Examples of such evaluations are presented in Fig.~\ref{fig:shifts}. The notable exceptions were  the blackbody radiation (BBR) shift, gravitational red shift and post-processed corrections between { UTC(AOS), UTC and TT (the SI second on the geoid).} 

{
The frequency stability (within 1~MHz) of the lattice 813~nm light was assured by narrowing and locking the lattice laser (pre-tuned to the magic wavelength  
368~554.58(28)~GHz\cite{Akatsuka10} with accuracy of 200~MHz by a HighFinesse WS6/200 wavemeter) to the ultrastable 689~nm laser by a Fabry-Perot transfer cavity. The length of the transfer cavity, i.e. wavelength of the cavity modes, was controlled by a piezoelectric transducer and temperature of the cavity spacer.
The lattice light shift  and its uncertainty are evaluated by  making a series of several interlaced locks corresponding to different lattice depths.
Different waists of the lattices in Sr1 and Sr2 (152 and 108~$\mu$m, respectively) and different depths of the lattices result in different values of measured residual light shift in Sr1 and Sr2.
}

 {The shift induced by the BBR  can be described as a static shift with a small dynamic correction \cite{middelmann1}. The static contribution is proportional to the differential static polarisability of the two clock states 
 \cite{middelmann2,marianna} and the mean square value of the electric field at temperature $T$. 
The dynamic correction is calculated similarly as in Ref.~\cite{middelmann2}.
 Two transitions to the states $5s5p$ $^3P_1$ and $5s5p$ $^1P_1$ contribute to the dynamic shift of the $5s^2$ $^1S_0$ ground state and four transitions to the states  $5s4d$ $^3D_1$, $5s6s$ $^3S_1$, $5p^2$ $^3P_1$ and $5s5d$ $^3D_1$ contribute to the dynamic shift  of the $5s5p$ $^3P_0$ excited state (see  Ref. \cite{marianna,porsev2}). 
 The relevant parameters of the transitions are taken from Ref. \cite{marianna,chem}. }

\begin{figure}[h!]
\centering
\includegraphics[width=0.6\textwidth]{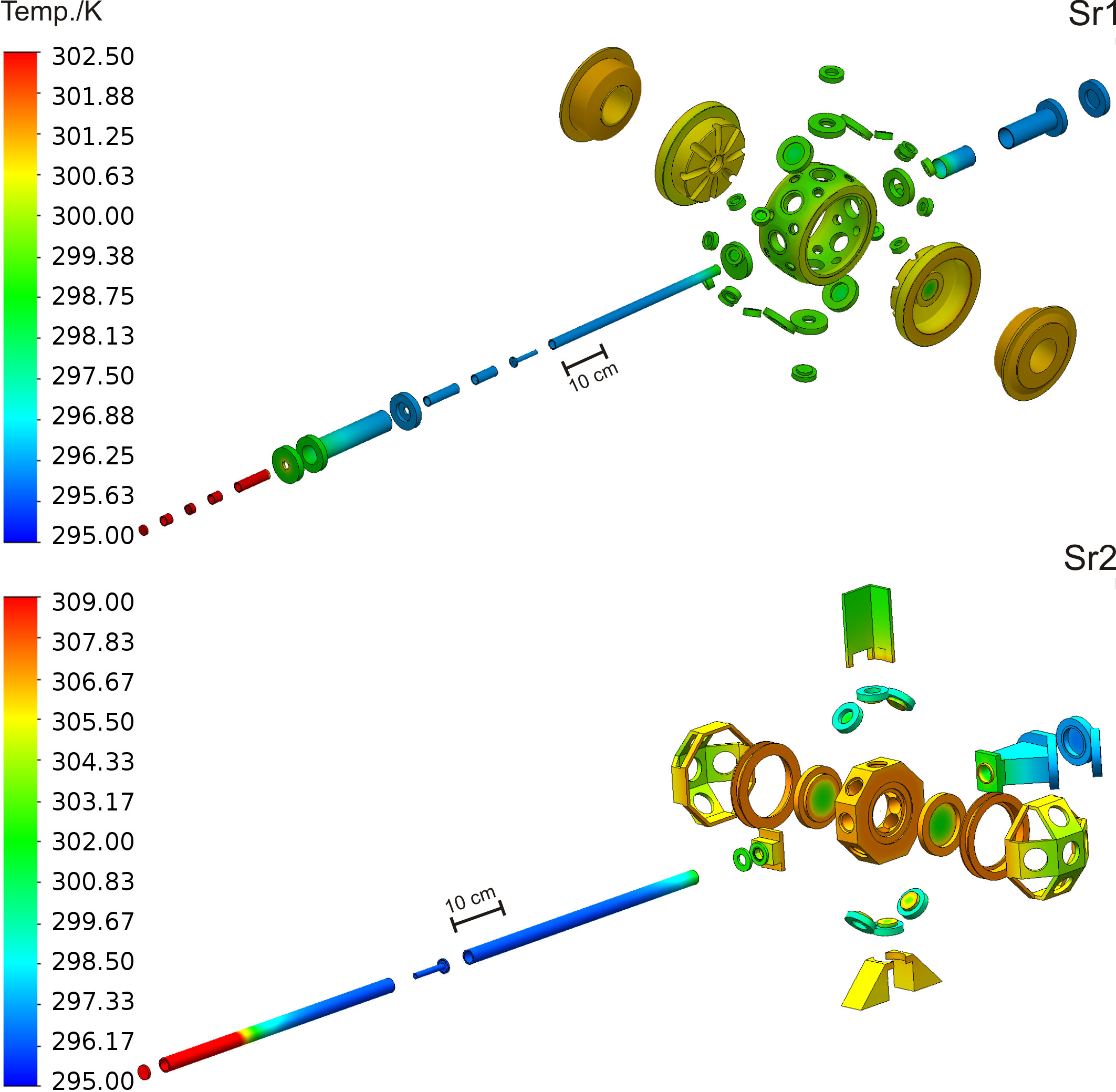}
\caption{  {{\bf The simulated temperature distribution of the vacuum systems and theirs surroundings.} The temperature of crucial points of the vacuum system is monitored during the experiment cycle by calibrated thermistors. The acquired data  are used to calculate the temperature distribution of the system by a finite-elements stationary thermal simulation.}  
Note that for the sake of clarity the temperature of the strontium ovens (above 770~K) is not included in the temperature legend.
 \label{fig:BBR}}
\end{figure}

 {The temperature of crucial points of the vacuum system is monitored during the experiment cycle by calibrated thermistors. The acquired data and an accurate model of the vacuum systems and theirs surroundings are used to  simulate the temperature distribution of the system (see Fig.~\ref{fig:BBR}).
In the simulation (based on finite-element method), the atoms are represented by a small vapour ball inside the vacuum chamber. The temperature probed by this ball is used to calculate the BBR experienced by the atoms.
The uncertainty of the shift is evaluated from calculations of the BBR for the maximum and minimum temperatures measured in the experiment.}

The UTC(AOS) signal in AOS in Borowiec is corrected with respect to the Earth's Geoid, therefore the measurements at KL FAMO also have to be  corrected with respect to the Geoid with the gravitational red shift. The local height over the Geoid, 50(2)~m, and the gravimetrically measured local value of the gravitational acceleration, 9.8127208(26)~m/s${}^2$, were used for this correction.

The uncertainty of the realization of the frequency of the UTC(AOS), i.e. the difference UTC - UTC(AOS), is estimated by comparing the UTC(AOS) and UTC(PTB) by the  Two Way Satellite  Frequency  and Time Transfer (TWSTFT) and by the proprietary GNSS time transfer system (TTS-4). The values of { UTC(AOS) - UTC and UTC - TT as well as  uncertainties of UTC(AOS)-UTC(PTB), UTC(AOS)-UTC and  UTC - TT }are reported in the Circular-T~\cite{Lewandowski06,CircT}.

The last evaluated uncertainty represents the finite resolution of the direct digital synthesizers (DDSs) driving the AOMs in the frequency chain of the clock lasers.

\begin{table}[h]
\caption{Accuracy budget for typical experimental conditions used in the measurement of the absolute frequency. All numbers are~in~hertz.  }
\label{tab:acc}
\centering
\begin{tabular}{ccc}
           \hline\hline
            Effects & \multicolumn{2}{c}{Shift(Uncert.)}  \\ 
             ~& Sr1 & Sr2\\
            \hline\hline
            Quadratic Zeeman & -151.9(1.7)& -115.42(2.7) \\ 
            Probe light & -3.82(0.35) & -9.02(0.37) \\ 
            Lattice light & -0.34(0.47) & { -1.55(0.48)}\\
            Collisions & 0.35(0.52) & 0.33(0.46) \\ 
            Blackbody radiation & -2.210(0.075) & -2.405(0.075) \\ 
            Grav. red shift & 2.34(0.10) &  2.34(0.10) \\
            UTC(AOS) -- UTC & {-0.40(0.43)} & {-0.40(0.43)} \\
            { UTC -- TT} & {0.10(0.11)} & {0.10(0.11)} \\
            DDS \& electronics & 0.00(0.16) & 0.00(0.12) \\ 
            \hline\hline
            {\bf Total:} & {\bf -155.9(1.9)} & {\bf -126.0(2.8)}\\
            \hline\hline
            \end{tabular} 
\end{table}

\begin{figure}[h!]
\centering
\includegraphics[width=\textwidth]{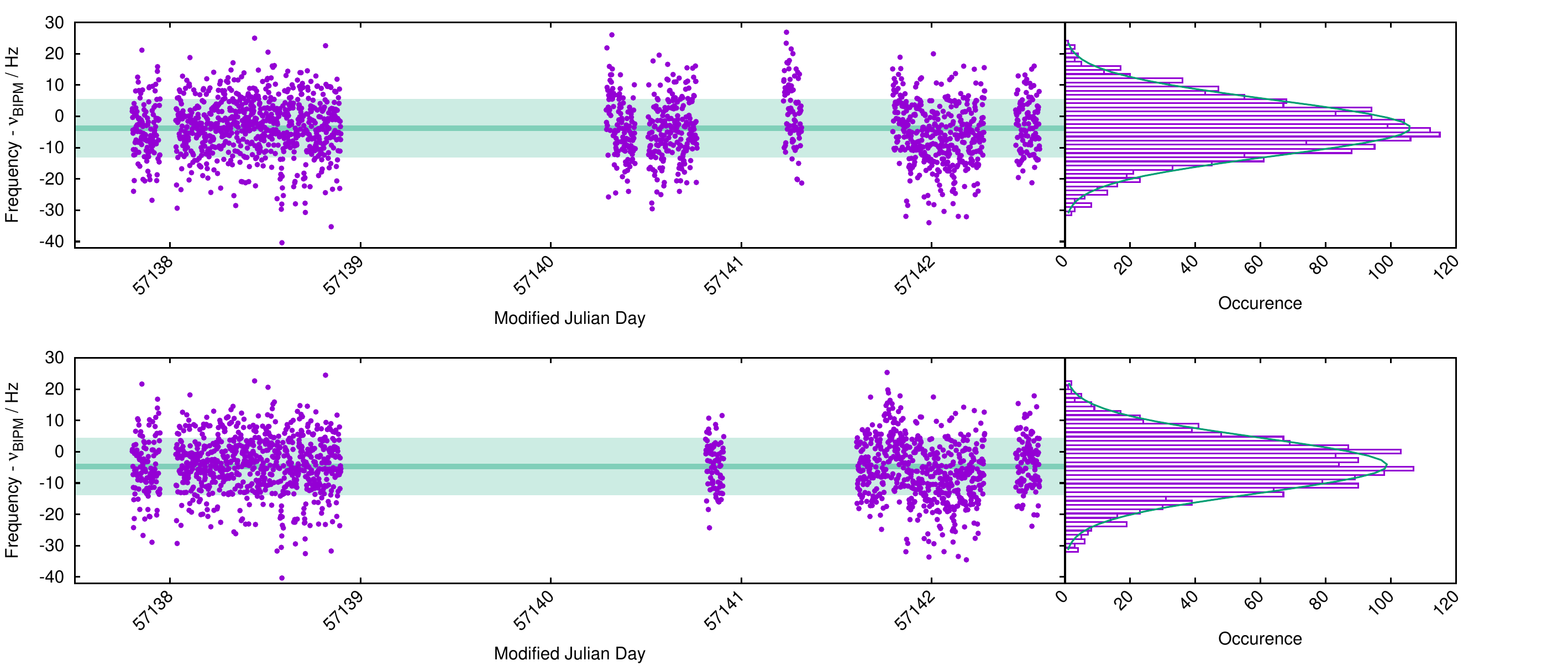}
\caption{{ {\bf  Frequencies of the ${}^{1}S_{0}$ -- ${}^{3}P_{0}$ clock transition in bosonic ${}^{88}$Sr recorded in Sr1 and Sr2 at the indicated MJD (top and bottom panels, respectively).} In the left panels each solid circle represent 100~s of averaged data, the light and dark-green regions represent 1~$\sigma$ standard deviation and standard deviation of the mean, respectively.  The  offset frequency $\nu_{BIPM}$ is  the  BIPM recommended frequency value\cite{BIPM88}. The right panels show a histogram of the frequency measurements with fitted Gaussian function.}
 \label{fig:evol}}
\end{figure}

The  described procedure yielded the absolute frequency of the  ${}^{1}S_{0}$ -- ${}^{3}P_{0}$ clock transition in bosonic ${}^{88}$Sr  equal to 429~228~066 418~008.3(1.9)${}_{syst}$(0.9)${}_{stat}$~Hz for Sr1 and  429~228~066 418~007.3(2.8)${}_{syst}$(0.9)${}_{stat}$~Hz for Sr2. 
Fig~\ref{fig:evol} documents the measurement record of  both Sr1 and Sr2 standards, binned at 100~s and histograms of the frequency measurements, plotted with an offset frequency $\nu_{BIPM}$=429~228~066 418~012~Hz, i.e the BIPM recommended value\cite{BIPM88}.

\section*{Discussion}

\begin{figure}[h!]
\centering
\includegraphics[width=0.5\columnwidth]{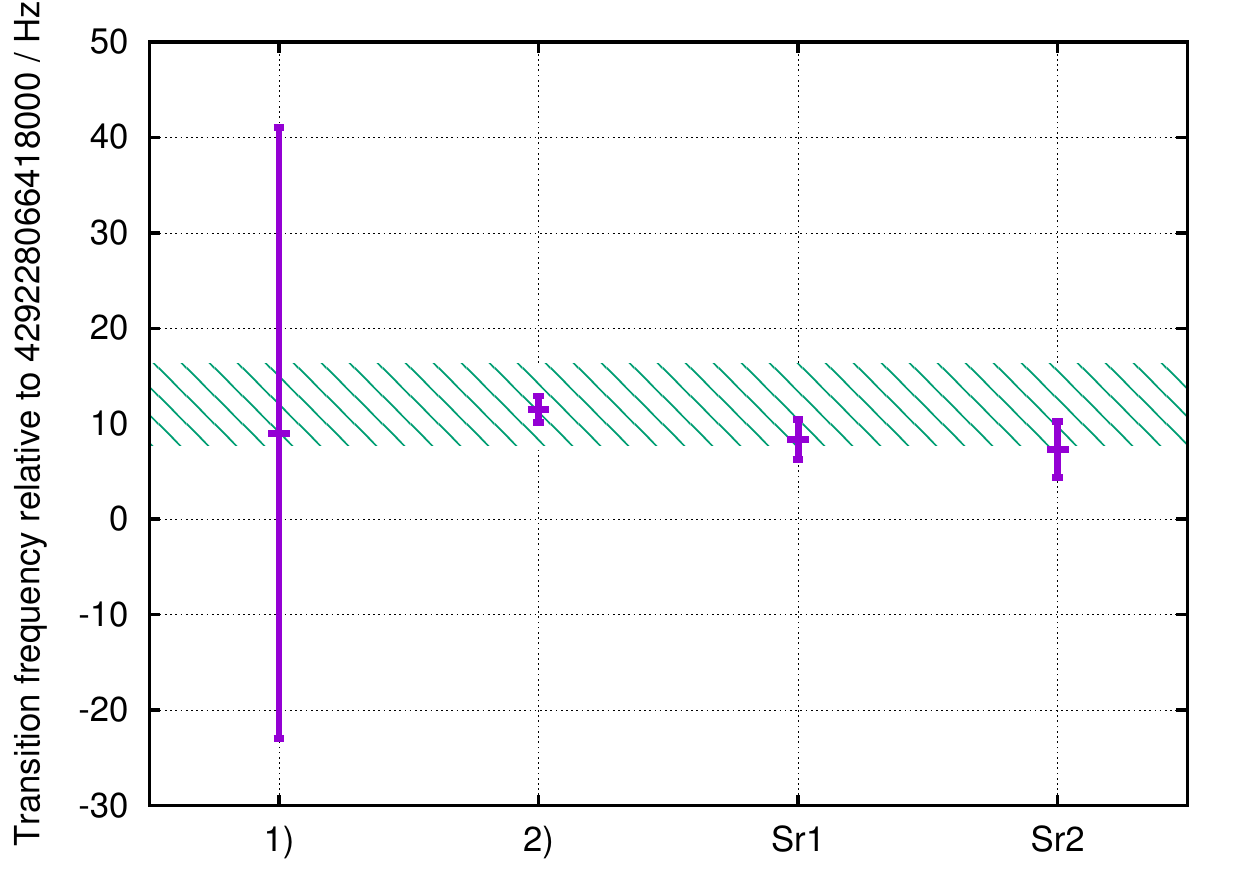}
\caption{ {\bf  Comparison of the ${}^{88}$Sr ${}^{1}S_{0}$ -- ${}^{3}P_{0}$ transition frequency with the previously known values.} 1) is the value directly measured in Ref. \cite{Baillard07}, 2) has been calculated from the frequency of the clock transition in ${}^{87}$Sr \cite{LeTargat13, Hinkley13, Falke14} and the measured isotope shift ${}^{88}$Sr-${}^{87}$Sr in Ref.~\cite{Akatsuka08}. Dashed horizontal band represents the value recommended by the BIPM~\cite{BIPM88}.
 \label{fig:zestawienie}}
\end{figure}

In Fig.~\ref{fig:zestawienie} we present comparison of the ${}^{88}$Sr ${}^{1}S_{0}$ -- ${}^{3}P_{0}$ transition frequency with the previously known values. The only direct measurement with ${}^{88}$Sr we found in the literature  has the uncertainty ten times bigger than the values reported in this paper~\cite{Baillard07}. The most precise value of the transition frequency was evaluated based on the measurement of the  isotope shift ${}^{88}$Sr-${}^{87}$Sr in Ref.~\cite{Akatsuka08} and from the frequency of the clock transition in ${}^{87}$Sr \cite{LeTargat13, Hinkley13, Falke14}. Dashed horizontal band in Fig.~\ref{fig:zestawienie} represents the value recommended by the BIPM~\cite{BIPM88}. We believe that better control of the magnetic field  would enable measurement of the  ${}^{88}$Sr ${}^{1}S_{0}$ -- ${}^{3}P_{0}$ transition frequency with accuracy at least order of magnitude better and recommendation of this transition as  SRS. 

\section*{Conclusion}

We have performed a series of measurements of the absolute frequency  of the ${}^{1}S_{0}$ -- ${}^{3}P_{0}$  transition in neutral ${}^{88}$Sr. The measurements has been made in two independent optical lattice clocks with an optical frequency comb referenced to the UTC(AOS) by a long distance stabilized fibre optic link.
Our results have comparable accuracy to those indirectly derived in Ref.~\cite{Akatsuka08} and one order of magnitude better accuracy than value measured directly and reported in Ref.~\cite{Baillard07}. Presented results agree with the recommendation of Bureau International des Poids et Mesures and should improve the accuracy of future recommendation. In conclusion, ${}^{1}S_{0}$ -- ${}^{3}P_{0}$ transition in the bosonic strontium seems to be a good candidate for practical representation of the second with stability of the order of 10${}^{-17}$, particularly for transportable systems.


\section*{Acknowledgement}

This work has been performed in the National Laboratory FAMO in Toru\'n and supported by the subsidy of  the Ministry of Science and Higher  Education.

Individual contributors were partially supported by
the Polish National Science Centre Projects No. 2012/07/B/ST2/00235, DEC-2013/11/D/ST2/02663,  DEC-2012/05/D/ST2/01914 and 2012/07/B/ST2/00251 and by the Foundation for Polish
Science TEAM Project co-financed by the EU within the European Regional Development Fund.

The  fiber optic link development and installation was supported by the Polish National Centre for Research and Development under the decision PBS1/A3/13/2012.

\section*{Author contributions statement}
MZ, PMo, MB: developed the concept and designed experiments, JZ, CR, WG, RC: conceived the optical clocks experiment, PMo, MB, MG: performed experiments, MZ, PMo, MB, RC: analysed data, MB, DBB: simulated the BBR shift, MB, PMo, MZ, DL, AC, PMa, WG, JZ, CR, RC : contributed in construction of the optical clocks apparatus, PK, \L{}\'S, ML: maintained the fibre link KL FAMO- AOS, JN, PD: realised and provided frequency referenced to UTC, MZ, WG: prepared the manuscript, WG, PMa, RC: technical support and conceptual advice.

\section*{Competing financial interests}

The authors declare no competing financial interests.

\end{document}